\documentclass[twocolumn]{jpsj2}
\usepackage{dcolumn}
\newcommand{\bsigma}{\mbox{\boldmath $\sigma$}}
\newcommand{\bxi}{\mbox{\boldmath $\xi$}}
\newcommand{\beeta}{\mbox{\boldmath $\eta$}}
\newcommand{\btau}{\mbox{\boldmath $\tau$}}
 
\newcommand{\bH}{\mbox{\boldmath $H$}} 
\newcommand{\sumdash}{\mathop{\sum{}^{'}}}
\newcommand{\proddash}{\mathop{\prod{}^{'}}}

\title{
Tensor Product State Formulation for the Spin 1/2 Antiferromagnetic XXZ Model on the Checkerboard Lattice
}

\author{
Nobuya \textsc{Maeshima}${}^{1,2}$\thanks{E-mail: maeshima@cp.cmc.osaka-u.ac.jp}
}

\inst{%
${}^1$ Large-scale Computational Science Division, Cybermedia Center, Osaka University\\ 1-32 Machikaneyamacho,
Toyonaka Osaka 560-0043 \\

${}^2$ Department of Engineering, Osaka Electro-Communication University, Neyagawa Osaka 572-8530}

\recdate{\today}

\abst{%
Vertical density matrix algorithm (VDMA), a tensor product state formulation of the ``higher-dimensional'' density matrix renormalization group, is applied to the spin 1/2 antiferromagnetic XXZ model on the checkerboard lattice. The VDMA was, in the preceding study, applied to the transverse field Ising model on the square lattice and the three-dimensional classical Ising model. In the present paper, its implementation procedure is modified in order to apply the VDMA to the XXZ model. Numerical accuracy of the VDMA is investigated for the XXZ model on the square lattice, which shows that the method gives reliable results for the ground state energy. In the frustrated region, VDMA results are compared with a simple calculation based on a magnetically disordered state. It is found that the weakly frustrated region is  in the N\'{e}el ordered phase, while in the strongly frustrated region the realized phase cannot be identified clearly by the obtained results.

}

\kword{%
DMRG, tensor product state, quantum spin system, frustration
}

\begin{document}
\sloppy
\maketitle

\section{Introduction}
\label{sec:intro}

The density matrix renormalization group (DMRG) method~\cite{orig-DMRG1,orig-DMRG2} has become  one of the standard numerical techniques for studying one-dimensional (1D) quantum systems and 2D classical systems
because of numerical accuracy and the ability to deal with large size systems.~\cite{dmrg-rev} The success of the DMRG has been stimulating us to extend the algorithm to the one that can handle higher-dimensional systems, principally 2D quantum systems and 3D classical systems.~\cite{LiPa,Nishino-Okunishi-CTTRG,Henelius,Martin,truly} Nevertheless, we have not achieved the definite success.

One of the key ideas for developing the ``higher-dimensional'' DMRG is the tensor product state (TPS).~\cite{njp,Niggemann1,Niggemann2,stripe} The TPS is a higher-dimensional generalization of the matrix product state (MPS) for 1D quantum spin systems.~\cite{Fannes}
On the basis of the idea of the TPS, several formulations for 3D classical systems have been proposed.~\cite{Okunishi-Nishino-KW,SCTPVA1,SCTPVA2,VDMA}
One of these TPS formulations is the vertical density matrix algorithm (VDMA).~\cite{VDMA}
In the preceding study, it is demonstrated that, besides the 3D classical Ising model, the 2D transverse field Ising (TFI) model, a most fundamental 2D quantum spin system, can be also dealt with by using VDMA.~\cite{VDMA} However, its implementation procedure for the 2D TFI model is completely equivalent to the one for the 3D classical Ising model because the 2D TFI model is exactly mapped to the 3D classical Ising model. Hence further development and modifications for general 2D quantum systems are desired for the progress of the higher-dimensional algorithm.

\begin{figure}
\includegraphics[width=4cm,clip]{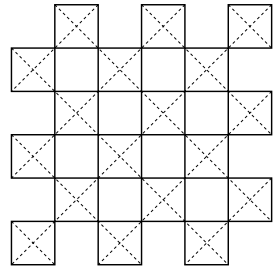}
\caption{The checkerboard lattice. Solid (dashed) lines denote the nearest (next-nearest) neighbor bonds.}
\label{fig:ppyro}
\end{figure}

In this paper, the VDMA is applied to the S=1/2 antiferromagnetic XXZ model on the checkerboard lattice. This lattice consists of the corner-sharing plaquettes that have the nearest neighbor and the next-nearest neighbor bonds, as depicted in Fig.~\ref{fig:ppyro}. Interestingly, in the isotropic case, this model is predicted to exhibit either the N\'{e}el ordered phase~\cite{kubokishi,Anderson,Rkubo,huse,reger,okabe,miyashita,barnes,weihong,sandvik,witte} or
 a magnetically disordered phase called the valence bond crystal (VBC).~\cite{palmer,canals,elhajal,moessner,fouet,brenig,brenig2,starykh,sind,auer}
 Hence it is appropriate to test the capability of the VDMA for both cases.


First, the VDMA calculation has been performed for the unfrustrated 2D XXZ model to compare the VDMA results with available quantum Monte Carlo (QMC) data. It is found that the VDMA gives quantitatively reliable results for the ground state energy; the relative deviations from the QMC results are the order of $1\%$.
In the frustrated region, the VDMA results are compared with a simple calculation based on the VBC.
It is found that in the weakly frustrated region the N\'{e}el ordered phase exists, while in the strongly frustrated region the realized phase cannot be identified by the obtained result; in the latter region, the VDMA result for the energy is greater than the upper bound of the energy obtained by using the VBC state.

This paper is organized as follows: In \S~\ref{sec:modifyvdma}, the definition of the model is given, and then the VDMA for the 2D XXZ model on the checkerboard lattice is explained. Section~\ref{sec:2DXXZresults} shows the numerical results for the unfrustrated case and the frustrated case. The last section is devoted to the conclusion.


\section{The Vertical Density Matrix Algorithm for the 2D XXZ Model}
\label{sec:modifyvdma}

\begin{figure}
\includegraphics[width=6cm,clip]{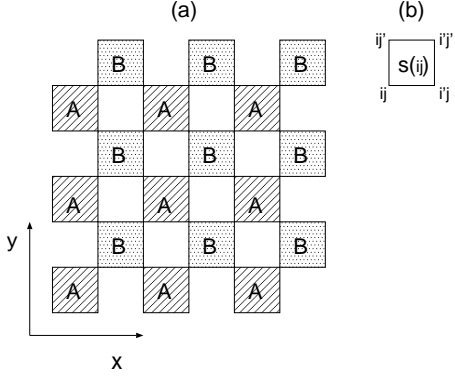}
\caption{(a): the checkerboard lattice with plaquettes divided
into two groups A and B. (b): the plaquette labelled $s(ij)$.}
\label{fig:check}
\end{figure}

\begin{figure}
\includegraphics[width=6cm,clip]{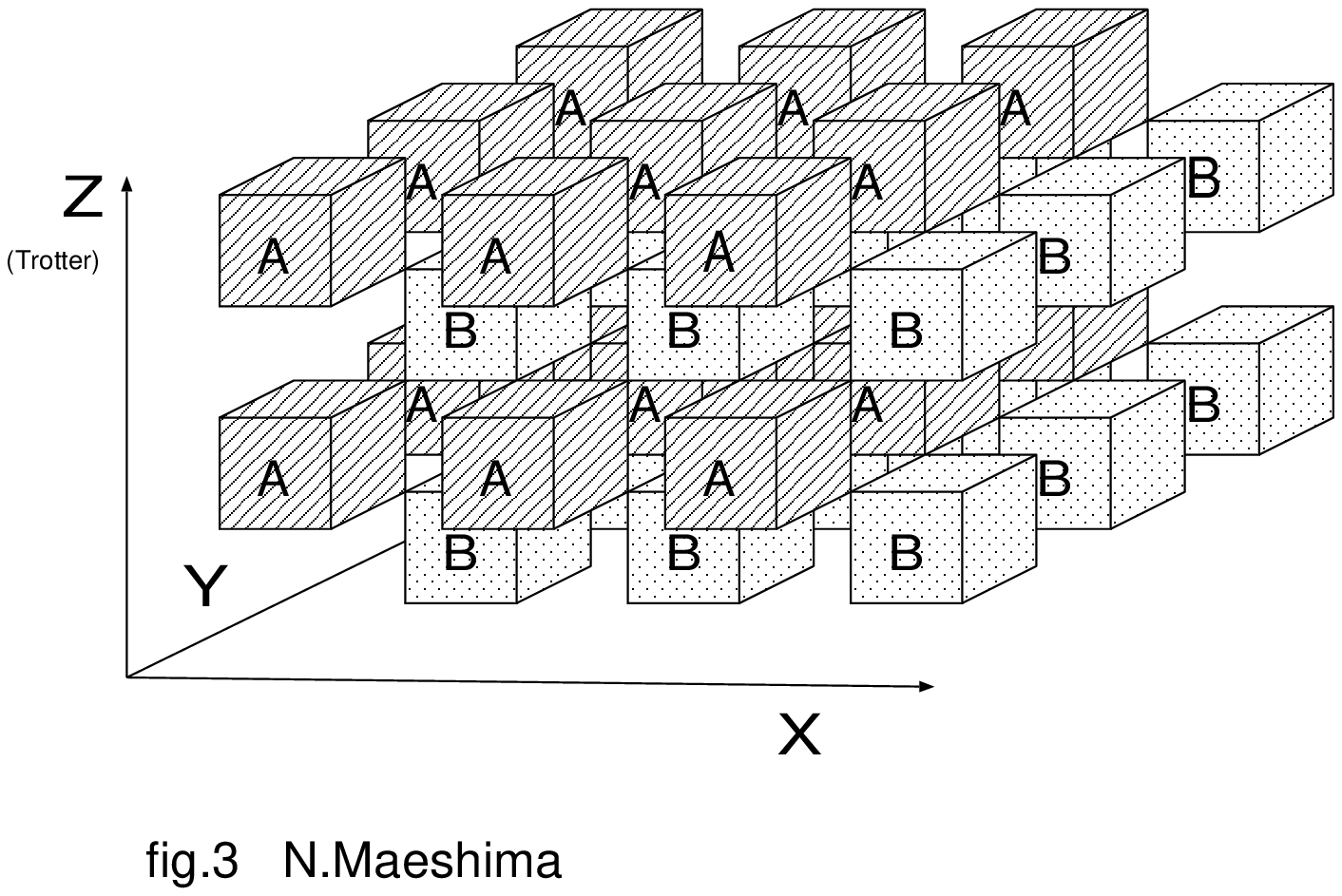}
\caption{The lattice structure of the 3D classical model mapped from the 2D XXZ model.
The $Z$-axis corresponds to the Trotter axis.}
\label{fig:3dcheck}
\end{figure}


The outline of the VDMA for 2D quantum spin systems is given in ref.17; after the mapping of 2D quantum spin systems to 3D classical spin systems by the Suzuki-Trotter decomposition,~\cite{suzuki} the VDMA is used to the 3D classical systems.
However, the implementation procedure for the 2D XXZ model is significantly different from the one for the 2D TFI model, which is described in ref.17, because of the difference in the lattice structure of the mapped 3D classical systems.

Let $\cal H$ be the Hamiltonian of the 2D quantum spin model under consideration; its explicit form is given as
\begin{equation}
 {\cal H} = \sum_{\langle (ij),(\bar{i}\bar{j}) \rangle}J_{(ij)(\bar{i}\bar{j})}h_{(ij)(\bar{i}\bar{j})}.
\label{eq:hamxxz}
\end{equation}
The sum runs over the nearest bonds and the next-nearest bonds shown in Fig.~\ref{fig:ppyro}, and  $i(j)$ is the $x(y)$ coordinate of a site. The local Hamiltonian $h_{(ij)(\bar{i}\bar{j})}$ with the anisotropy parameter $g(>$0) is defined as
\begin{equation}
h_{(ij)(\bar{i}\bar{j})} = g(S^x_{ij}S^x_{\bar{i}\bar{j}} + S^y_{ij}S^y_{\bar{i}\bar{j}}) + S^z_{ij}S^z_{\bar{i}\bar{j}},
\end{equation}
where $S^\alpha_{ij}$ is the $\alpha$-component ($\alpha=x,y,$ or $z$) of the spin operator at the site $ij$. The exchange coupling $J_{(ij)(\bar{i}\bar{j})}$ is set to unity for the nearest neighbor bonds and $j_d$ for the next nearest neighbor bonds.

The first step of the algorithm is to map the 2D quantum spin model into a 3D classical spin model.
We follow the standard ``checkerboard'' decomposition~\cite{loh} for 2D quantum systems.
Divide all the corner-sharing plaquettes in the checkerboard lattice into two groups A and B, as shown in Fig.~\ref{fig:check} (a). A plaquette that has 4 sites ($(ij)$, $(ij')$, $(i'j)$, and $(i'j')$) is labeled as $s(ij)$, where $i'=i+1$ and $j'=j+1$ (see Fig.~\ref{fig:check}(b)). Then the Hamiltonians of the two groups A and B  are written as
\begin{equation}
{\cal H}_{A} = \sum_{s(ij) \in A} \tilde{h}_{s(ij)},\quad{\rm and}\quad
{\cal H}_{B} = \sum_{s(ij) \in B} \tilde{h}_{s(ij)}.\label{eq:divideham}
\end{equation}
Here $\tilde{h}_{s(ij)}$ is the local Hamiltonian of the 4 spins on the plaquette $s(ij)$;
\begin{eqnarray}
\tilde{h}_{s(ij)} &=& h_{(ij)(i'j)} + h_{(i'j)(i'j')} + 
h_{(i'j')(ij')} + h_{(ij')(ij)} \nonumber \\ &+& j_d [h_{(ij)(i'j')}  + h_{(i'j)(ij')}].
\end{eqnarray}

Then the following Suzuki-Trotter transformation
\begin{equation}
Z = \lim_{L \to \infty} {\rm Tr}(e^{-\epsilon {\cal H_A}} e^{-\epsilon {\cal H_B}})^L
\end{equation}
is used to map the 2D XXZ model at temperature $T$ into the classical model on the 3D lattice
with $2L$ Trotter size (see Fig.~\ref{fig:3dcheck}).
It is found that the mapped 3D lattice is composed of the corner-sharing cubes, each of which have 
8 Ising spins at its corners. The Boltzmann weight for an unit cube is represented as
\begin{eqnarray}
W\left( \begin{array}{c}
\bar{\bsigma}_{ij} \\
\bsigma_{ij} \\ \end{array} \right)
\equiv  \langle\bar{\bsigma}_{ij}|\exp( \epsilon \tilde{h}_{s(ij)})|
\bsigma_{ij}\rangle,
\label{eq:defBWxxz}
\end{eqnarray}
where we use the notation $\epsilon=J/(Lk_{\rm B}T)$ and
$|\bsigma_{ij}\rangle = |\sigma_{ij}\rangle|\sigma_{i'j}\rangle|\sigma_{i'j'}\rangle|\sigma_{ij'}\rangle$. $|\sigma_{ij}\rangle$ is the $S^z$-representation of the spin at the  site $ij$.

The next step is to take the $L \to \infty$ limit with fixed $\epsilon$, which is equivalent to taking the zero temperature limit of the original 2D model. Then the partition function $Z$
can be expressed as a simple form.
Consider $e^{-\epsilon {\cal H}_A} e^{-\epsilon {\cal H}_B}$ as a matrix $T$, and
let $E_{\rm max}$ and $|\psi^{r(l)}_{\rm max}\rangle$ be the maximum eigenvalue and the corresponding right (left) eigenstate of $T$.
Using the power method, we can calculate the partition function as follows;
\begin{eqnarray}
 Z &=& \lim_{L \to \infty, \rm fixed \ \epsilon }{\rm Tr} \ T^L \nonumber \\
   &=& \lim_{L \to \infty, \rm fixed \ \epsilon } E_{\rm max}^L \langle\psi^l_{\rm max}|\psi^r_{\rm max}\rangle.
\end{eqnarray}
Hence it is found that $Z$ is equivalent to the norm  $\langle\psi^l_{\rm max}|\psi^r_{\rm max}\rangle$ except for the trivial coefficient $E_{\rm max}^L$. Here it should be stressed that we now need both $|\psi^r_{\rm max}\rangle$ and $\langle\psi^l_{\rm max}|$ to evaluate the partition function because $T$ is an asymmetric matrix. In contrast, for the 2D TFI model we can use a symmetric matrix, and therefore we needs only the right (or left) eigenstate.~\cite{VDMA}

In the framework of the VDMA, each eigenstate is given in a form of the TPS. 
Let us consider the TPS structure of $|\psi_{\rm max}^r\rangle$ using the power method.
Define a vector $|\psi^r_{L}\rangle$ as
\begin{equation}
|\psi_{L}^r\rangle = T^L|\psi_0\rangle,
\end{equation}
where $|\psi_{0}\rangle$ is an ``initial'' vector that is not orthogonal to $|\psi^r_{\rm max}\rangle$.
Here we use $|\psi_{0}\rangle$ given by a TPS;
\begin{equation}
|\psi\rangle = \prod_{s(ij)\in B} w(\btau^0_{ij})|\btau^0_{ij}\rangle,
\label{eq:initialt}
\end{equation}
where $w(\btau^0_{ij})$ is a local tensor.
Then $|\psi^r_{L}\rangle$ is rewritten as the TPS;
\begin{eqnarray}
\langle[\bsigma^{2L}]|\psi_{L}^r\rangle &=& \sum_{\bxi_{ij}: all\,ij} \, 
\left[ \prod_{s(ij) \in A}
X^A_L\left( \begin{array}{c} \bsigma^{2L}_{ij} \\
\bxi^{2L}_{ij} \\ \end{array} \right) \right] \nonumber \\ &\times&
\left[ \prod_{s(ij) \in B} 
X^B_L\left( \bxi^{2L}_{ij} \right) \right] \label{eq:tpsright},
\end{eqnarray}
where 
\begin{equation}
|[\bsigma^{2L}]\rangle \equiv \prod_{s(ij) \in A} |\bsigma^{2L}_{ij}\rangle.
\end{equation}
The two tensors $X^A$ and $X^B$ are defined as
\begin{eqnarray}
X^A \left( \begin{array}{c} \bsigma^{2L}_{ij} \\ \bxi^{2L}_{ij} \\ \end{array} \right) &=&
W \left( \begin{array}{l} \bsigma^{2L}_{ij} \\ \bsigma^{2L-1}_{ij} \\ \end{array} \right)
W \left( \begin{array}{c} \bsigma^{2L-2}_{ij} \\ \bsigma^{2L-3}_{ij} \\ \end{array} \right)
\cdots \nonumber \\ &\times&
W \left( \begin{array}{c} \bsigma^{4}_{ij} \\ \bsigma^{3}_{ij} \\ \end{array} \right)
W \left( \begin{array}{c} \bsigma^{2}_{ij} \\ \bsigma^{1}_{ij} \\ \end{array} \right), 
\end{eqnarray}
and
\begin{eqnarray}
X^B \left( \bxi^{2L}_{ij} \right) &=& \sum_{\btau^0_{ij}}
W \left( \begin{array}{c} \bsigma^{2L-1}_{ij} \\ \bsigma^{2L-2}_{ij} \\ \end{array} \right)
W \left( \begin{array}{c} \bsigma^{2L-3}_{ij} \\ \bsigma^{2L-4}_{ij} \\ \end{array} \right)
\cdots \nonumber \\ &\times&
W \left( \begin{array}{c} \bsigma^{3}_{ij} \\ \bsigma^{2}_{ij} \\ \end{array} \right)
W \left( \begin{array}{c} \bsigma^{1}_{ij} \\ \btau^{0}_{ij} \\ \end{array} \right)
w(\btau^0_{ij}),
\end{eqnarray}
where
$\bxi^{2L}_{ij}=(\bsigma^1_{ij}\bsigma^2_{ij}\cdots\bsigma^{2L-1}_{ij})$ is an auxiliary
variable.~\cite{VDMA}

\begin{figure}
\includegraphics[width=7cm,clip]{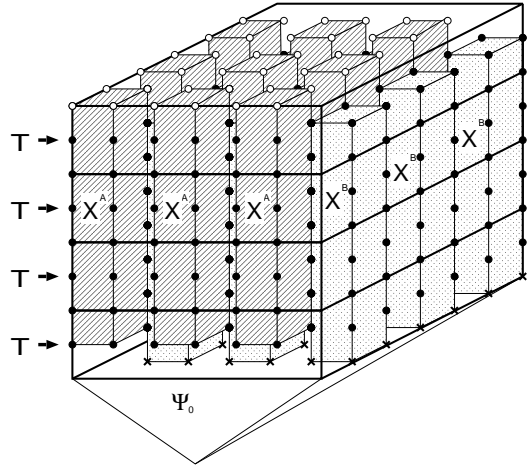}
\caption{The graphical representation of the vector $|\psi^r_{L}\rangle$ (L=4).
Filled circles and crosses show the summed spin variables in eq.~(\ref{eq:tpsright}).
Open circles represent the fixed spin variables.
}
\label{fig:phir}
\end{figure}

\begin{figure}
\includegraphics[width=7cm,clip]{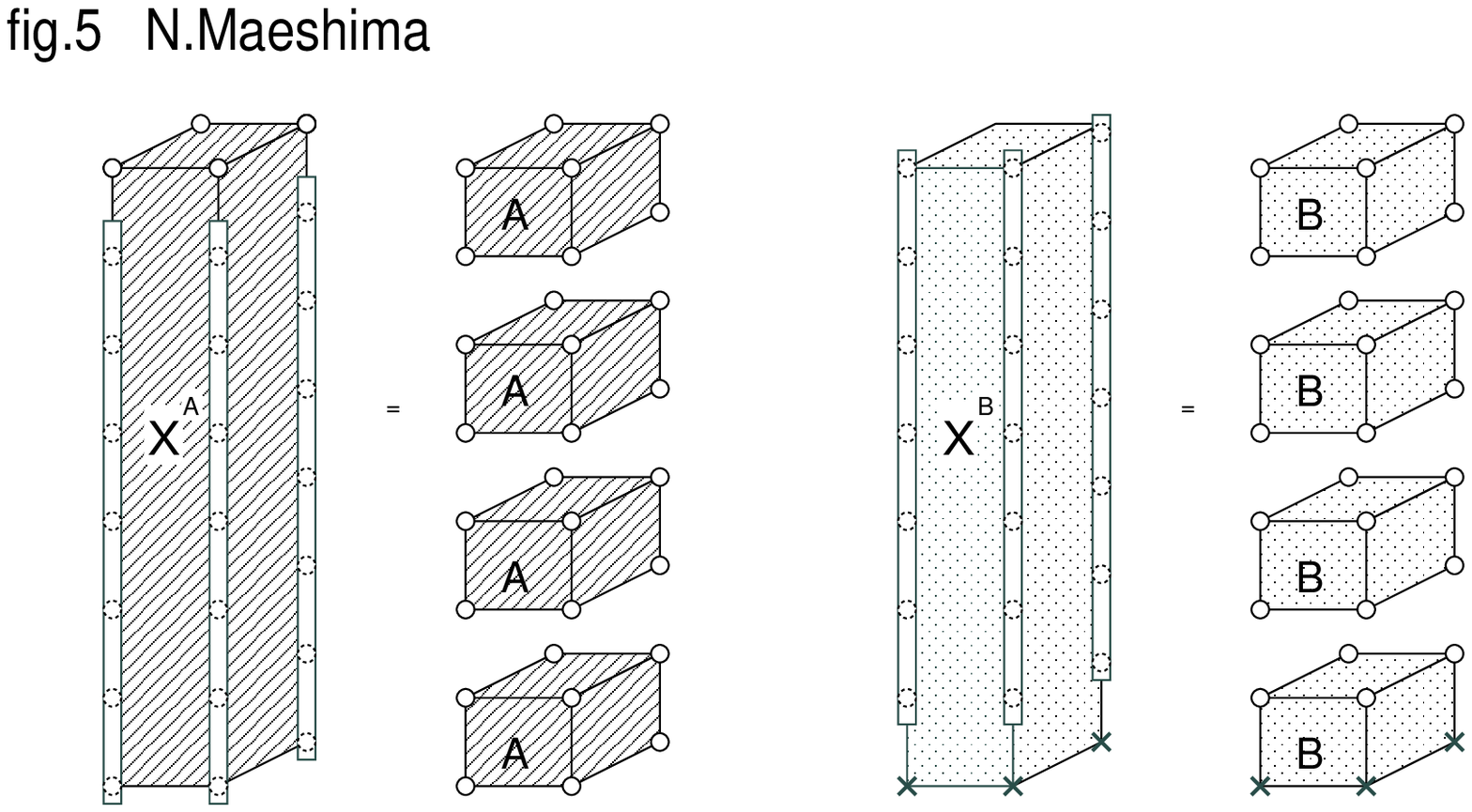}
\caption{The two tensors $X^A$ and $X^B$.}
\label{fig:tensor}
\end{figure}

It should be noted that $|\psi^r_{L}\rangle$ consists of two types of  tensors $X^A$ and $X^B$
whereas only one tensor is needed for the 2D TFI model.~\cite{VDMA}
This difference stems from the structure of $|\psi^r_{L}\rangle$, which is shown in Fig.~\ref{fig:phir}. We can see that $|\psi^r_{L}\rangle$ is constructed by  $X^A$ and $X^B$, which are represented by rectangular parallelepipeds composed of the cubes vertically piling up (see Fig.~\ref{fig:tensor}). 
We also found that the bare spin variables $\{\sigma^{2L}_{ij}\}$ are on the upper faces of the rectangular parallelepipeds corresponding to the tensor $X^A$, not to $X^B$. Hence the tensor $X^A$ has the bare spin variables $\bsigma^{2L}_{ij}$ as indices and $X^B$ does not so.

By taking the limit $L \to \infty$, the right maximum-eigenvalue eigenvector $|\psi^r_{\rm max}\rangle$ is obtained;
\begin{equation}
|\psi^r_{\rm max}\rangle = {\rm const} \times \lim_{L \to \infty} |\psi^r_L\rangle.
\end{equation}
Therefore $|\psi^r_{\rm max}\rangle$ is represented as the TPS composed of the tensors $X^A$ and $X^B$ with the auxiliary variables $\bxi^{\infty}_{ij}$.
The left maximum-eigenvalue eigenvector $\langle \psi^l_{\rm max}|$ is also
represented as the same TPS form with the ``transposed local tensor'',
\begin{eqnarray}
\bar{X}^A \left( \beeta^{2L}_{ij} \right) &=& \sum_{\bar{\btau}^0_{ij}} w(\bar{\btau}^0_{ij})
W \left( \begin{array}{c} \bar{\btau}^0_{ij} \\ \bar{\bsigma}^{1}_{ij} \\ \end{array} \right)
W \left( \begin{array}{c} \bar{\bsigma}^{2}_{ij} \\ \bar{\bsigma}^{3}_{ij} \\ \end{array} \right)
\cdots \nonumber \\ &\times&
W \left( \begin{array}{c} \bar{\bsigma}^{2L-4}_{ij} \\ \bar{\bsigma}^{2L-3}_{ij} \\ \end{array} \right)
W \left( \begin{array}{c} \bar{\bsigma}^{2L-2}_{ij} \\ \bar{\bsigma}^{2L-1}_{ij} \\ \end{array} \right), 
\end{eqnarray}
and
\begin{eqnarray}
& &\bar{X}^B\left( \begin{array}{c} \beeta^{2L}_{ij}  \\ \bsigma^{2L}_{ij} \\ \end{array} \right) \nonumber \\ &=&
W \left( \begin{array}{c} \bar{\bsigma}^{1}_{ij} \\ \bar{\bsigma}^{2}_{ij} \\ \end{array} \right)
W \left( \begin{array}{c} \bar{\bsigma}^{3}_{ij} \\ \bar{\bsigma}^{4}_{ij} \\ \end{array} \right)
\cdots \nonumber  \\ &\times&
W \left( \begin{array}{c} \bar{\bsigma}^{2L-3}_{ij} \\ \bar{\bsigma}^{2L-2}_{ij} \\ \end{array} \right)
W \left( \begin{array}{l} \bar{\bsigma}^{2L-1}_{ij} \\ \bsigma^{2L}_{ij} \\ \end{array} \right),
\end{eqnarray}
where $\beeta^{2L}_{ij}\equiv(\bar{\bsigma}^1_{ij}\bar{\bsigma}^2_{ij}\cdots\bar{\bsigma}^{2L-1}_{ij})$.
When we define the vector
\begin{equation}
\langle \psi^l_{L}| = \langle\psi_0|T^L,
\end{equation}
$\langle \psi^l_{\rm max}|$ is represented as the following TPS form;
\begin{eqnarray}
& &\langle\psi_{\rm max}^l|[\bsigma^{\infty}]\rangle = {\rm const}\times \lim_{L \to \infty} \langle \psi^l_L|[\bsigma^{2L}]\rangle \\
&=&{\rm const} \times \lim_{L\to\infty}\sum_{\bxi_{ij}: all\,ij} \, 
\left[ \prod_{s(ij) \in A}
\bar{X}^A_L\left( \beeta^{2L}_{ij} \right) \right] \nonumber \\ &\times&
\left[ \prod_{s(ij) \in B} 
\bar{X}^B_L\left( \begin{array}{c} \beeta^{2L}_{ij}\\ \bsigma^{2L}_{ij} \end{array} \right) 
\right] \label{eq:tpsleft}.
\end{eqnarray}
It should be also noted that, from the definitions of the tensors, the following relations hold;
\begin{eqnarray}
X^A &=& \bar{X}^B \nonumber \\
X^B &=& \bar{X}^A \label{eq:equiv}.
\end{eqnarray}
These relations are used to obtain renormalized tensors, as is mentioned later.

\begin{figure}
\includegraphics[width=7cm,clip]{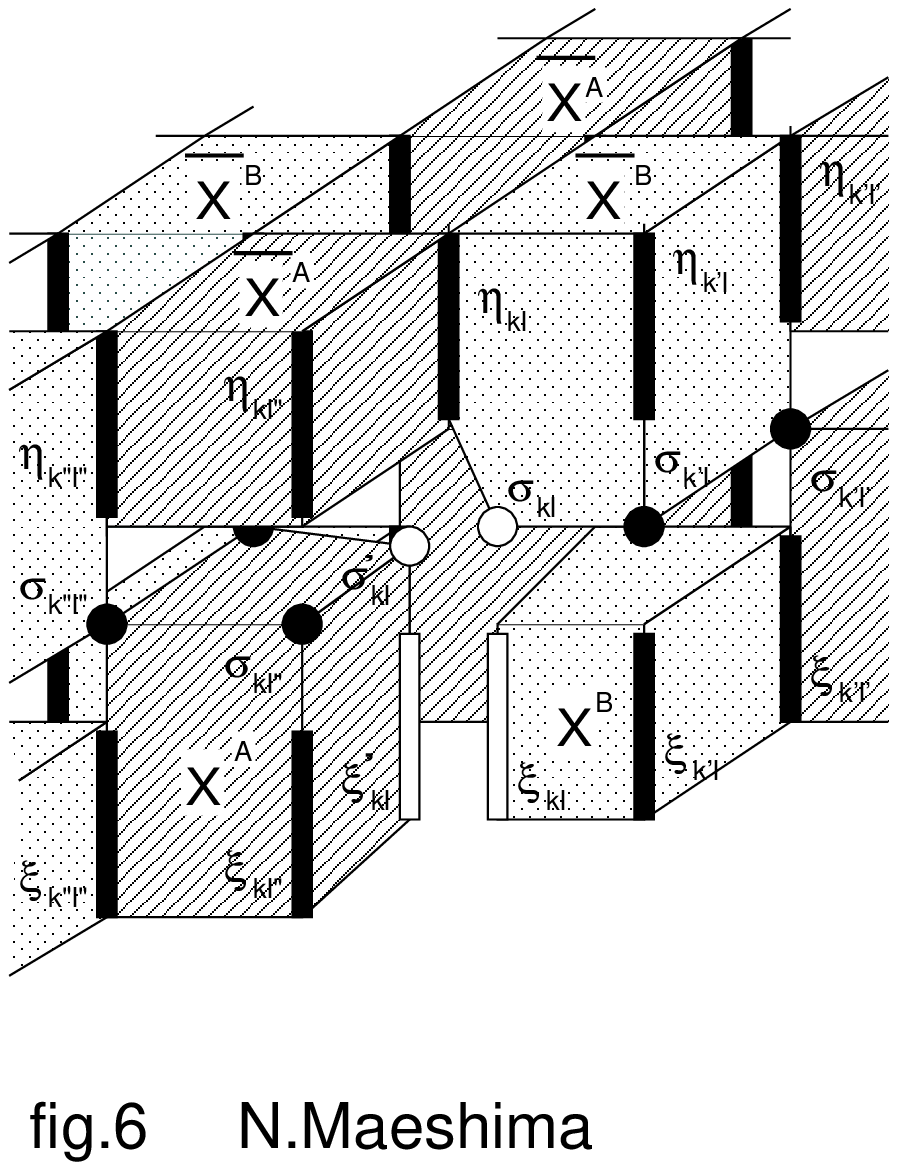}
\caption{The vertical density matrix of the 3D classical model mapped from the 2D XXZ model.
Open circles (rectangles) represent the fixed spin variables in eq.~(\ref{eq:getrhoxxz}), and filled circles (rectangles) represent the summed variables.
}
\label{fig:xxzvdm}
\end{figure}

The key point of the VDMA is to reduce the dimensions of the auxiliary variables
$\bxi$ and $\beeta$ using the vertical density matrix (VDM) $\rho$.
By taking account of the lattice structure of the 3D classical model,
we can define the VDM as
\begin{eqnarray}
& &\rho_{kl}(\sigma'_{kl} \xi'_{kl}|\sigma_{kl} \xi_{kl})= \nonumber \\ & &
\sumdash_{[\bsigma],[\bxi],[\beeta]}
\left[\proddash_{s(ij)\in A}
\bar{X}^A \left( \beeta_{ij} \right)
{X}^A \left( \begin{array}{c} \bsigma_{ij} \\ \bxi_{ij}\end{array} \right) \right] \nonumber \\ &\times&
\left[\proddash_{s(ij)\in B}
\bar{X}^B \left( \begin{array}{c}\beeta_{ij} \\ \bsigma_{ij} \end{array}  \right)
\bar{X}^B \left( \bxi_{ij} \right) \right]  
\bar{X}^A\left(\beeta_{k''l''} \right) \nonumber \\ &\times&
X^A\left( \begin{array}{c} \check{\bsigma}_{k''l''} \\ \check{\bxi}_{k''l''}\\ \end{array}\right)
\bar{X}^B\left( \begin{array}{c} \beeta_{kl}\\ \bsigma_{kl}\end{array} \right)
 X^B \left(\bxi_{kl}\right) \label{eq:getrhoxxz} ,
\end{eqnarray}
 where $\sum^{'}$ denotes the configuration sum for the all spin variables
except $\sigma_{kl}^{'},\xi_{kl}^{'},\sigma_{kl}$, and $\xi_{kl}$.
$\proddash_{s(ij)\in A}$ denotes the product for the plaquettes of group $A$ except
$s(k^{``},l^{``})\equiv s(k-1,l-1)$, and
$\proddash_{s(ij)\in B}$ is the product for the plaquettes of group $B$ except
$s(k,l)$. The checked spin is defined as
\begin{equation}
\left( \begin{array}{c} \check{\bsigma}_{k''l''} \\ \check{\bxi}_{k''l''}\\ \end{array}\right) =
\left( \begin{array}{cccc}
\sigma_{k''l''}&\sigma_{kl''}&\sigma^{'}_{kl}&\sigma_{k''l} \\
\xi_{k''l''}&\xi_{kl''}&\xi^{'}_{kl}&\xi_{k''l}\end{array} \right).
\end{equation}
The locations of the spin variables are shown in Fig.~\ref{fig:xxzvdm}

The diagonalization of the VDM $\rho$ gives us the projection operators to reduce the dimensions of $\bxi$ and $\beeta$.
Since the  VDM $\rho$ is asymmetric, we obtain right-eigenvectors $U^r(\sigma\xi|\tilde{\xi})$ and the left-eigenvectors $U^l(\sigma\xi|\tilde{\xi})$, with the corresponding eigenvalues $\omega_{\tilde{\xi}}$ in the decreasing order $\omega_1 \ge \omega_2 \ge \cdots$. In the actual implementation of the VDMA for the 2D XXZ model, all the eigenvalues $\omega_{\tilde{\xi}}$ are real and positive except for the first several steps of the iteration. By taking $U^r(\sigma\xi|\tilde{\xi})$ and $U^l(\sigma\xi|\tilde{\xi})$ with $\tilde{\xi} \in 1,\cdots ,M$, we construct the projection operators $U^r$ and $U^l$.

The renormalization procedure is described as follows. On the basis of the variational principle for the partition function,~\cite{nishino}
we use the operator $U^r$ to renormalize the spin variables on each ``edge'' of $X^B$, and 
the operator $U^l$ is used for $X^A$. Thus the renormalized tensors $\tilde{X}^A$ and $\tilde{X}^B$ are obtained as
\begin{eqnarray}
& &\tilde{X}^A \left( \tilde{\bxi}_{ij} \right) \nonumber \\ &\equiv&
\sum_{ \bsigma_{ij},\bxi_{ij} } 
X^A\left( \begin{array}{cccc}
\sigma_{ij}&\sigma_{i'j}&\sigma_{i'j'}&\sigma_{ij'} \\
\xi_{ij}&\xi_{i'j}&\xi_{i'j'}&\xi_{ij'}\end{array} \right) \nonumber \\ &\times&
U^l( \sigma_{ij} \xi_{ij}|\tilde{\xi}_{ij} ) 
U^l( \sigma_{i'j} \xi_{i'j}|\tilde{\xi}_{i'j} )
U^l( \sigma_{i'j'} \xi_{i'j'}|\tilde{\xi}_{i'j'} )   \nonumber \\ &\times&
U^l( \sigma_{ij'} \xi_{ij'}|\tilde{\xi}_{ij'} ) \label{eq:Renormxa}.
\end{eqnarray}
and
\begin{eqnarray}
& &\tilde{X}^B \left( \begin{array}{c} \btau_{ij} \\ \tilde{\bxi}_{ij} \end{array} \right)
\nonumber \\
&\equiv&
\sum_{ \bsigma_{ij},\bxi_{ij} } 
 W\left( \begin{array}{cccc}
\tau_{ij}&\tau_{i'j}&\tau_{i'j'}&\tau_{ij'} \\
\sigma_{ij}&\sigma_{i'j}&\sigma_{i'j'}&\sigma_{ij'}\end{array} \right) \nonumber \\ &\times&
X^B\left( \xi_{ij}\xi_{i'j}\xi_{i'j'}\xi_{ij'}\right) 
U^r( \sigma_{ij} \xi_{ij}|\tilde{\xi}_{ij} ) 
U^r( \sigma_{i'j} \xi_{i'j}|\tilde{\xi}_{i'j} )        \nonumber \\ &\times&
U^r( \sigma_{i'j'} \xi_{i'j'}|\tilde{\xi}_{i'j'} )  
U^r( \sigma_{ij'} \xi_{ij'}|\tilde{\xi}_{ij'} ) \label{eq:renormxb}.
\end{eqnarray}
The transposed tensor is obtained by using the relation eq.~(\ref{eq:equiv}).

Here, it should be noted that the forms of the tensors alternate by one renormalization procedure; the new tensor $\tilde{X}^{A}$ has only the auxiliary variables as the indices, whereas $\tilde{X}^B$ has both the bare spins and auxiliary variables. Therefore, $\tilde{X}^{A(B)}$ has the same form as $X^{B(A)}$, which suggests that we should interchange the tensors as
\begin{equation}
X^{A}_{new} = \tilde{X}^B,\quad {\rm and} \quad X^B_{new} = \tilde{X}^A,
\end{equation}
to obtain new tensors $X^A_{new}$ and $X^B_{new}$ for the next step.

The method of calculating the VDM $\rho$ is essentially the same as that for the 3D Ising model.~\cite{VDMA}
Define two tensors
\begin{eqnarray}
G^A\left(  \begin{array}{c}\beeta_{ij}\\ \bsigma_{ij}\\ \bxi_{ij} \end{array}  \right) \equiv
\bar{X}^A \left( \beeta_{ij} \right)
X^A \left( \begin{array}{c} \bsigma_{ij} \\ \bxi_{ij} \end{array} \right), \label{defGA}
\end{eqnarray}
and
\begin{eqnarray}
G^B\left(  \begin{array}{c}\beeta_{ij}\\ \bsigma_{ij}\\ \bxi_{ij}\end{array}  \right) \equiv
\bar{X}^B \left( \begin{array}{c} \beeta_{ij} \\ \bsigma_{ij} \end{array} \right)
X^B \left( \beeta_{ij} \right), \label{defGB}
\end{eqnarray}
and regard  $G^A$ and $G^B$ as the Boltzmann weights of a new 2D classical statistical
system. Then we can calculate the VDM and various observables
using the standard DMRG method for 2D classical statistical
systems~\cite{nishino} or the corner transfer matrix renormalization group (CTMRG)~\cite{ctmrg}.
It should be also noted that, in the DMRG for the 2D classical system (or in the CTMRG), the number of the retained basis is denoted by $m$, which is discriminated from the number of state $M$ of the auxiliary spin variables $\xi$ and $\eta$.

Finally some technical details of the VDMA are mentioned. Since the VDMA deals with infinite-size systems, we can calculate the spontaneous staggered magnetization directly by applying the finite staggered magnetic field. In actual calculation the staggered magnetic field is introduced as follows. Only at the initial step of the VDMA we use the Boltzmann weight defined as
\begin{eqnarray}
W_{initial}\left( \begin{array}{c}
\bar{\bsigma}_{ij} \\
\bsigma_{ij} \\ \end{array} \right)
\equiv  \langle\bar{\bsigma}_{ij}|\exp[ \epsilon \tilde{h}_{s(ij)}(\bH_s)]|
\bsigma_{ij}\rangle,
\label{eq:defBWxxzinit}
\end{eqnarray}
where the local Hamiltonian is
\begin{eqnarray}
\tilde{h}_{s(ij)}(\bH_s) &=& \tilde{h}_{s(ij)} - H^z_s(S^z_{ij}-S^z_{ij'}-S^z_{i'j}+S^z_{i'j'}) \nonumber \\
&-& H^x_s(S^x_{ij}-S^x_{ij'}-S^x_{i'j}+S^x_{i'j'}),
\end{eqnarray}
and $\bH_s=(H^x_s,H^y_s=0,H^z_s)$ is the staggered field.

How the staggered field is applied influences converged results and the speed of the convergence.
We should not apply the strong staggered field that conflicts the anisotropy of the system.
For example, if we set $H^x_s >> 1$ and $H^z_s=0$ for the Ising-like model ($g<1$), the convergence often becomes very slow because such an inappropriate field can lead the result to a quasi-stable XY-N\'{e}el state, not to the Ising-like N\'{e}el ground state.

The author also gives a comment on the initial tensor $w(\btau)$.
It is found that $w(\btau)$ does not have a strong influence to the converged results and the speed of the convergence if the initial TPS made from $w(\btau)$ is not orthogonal to $|\psi^r_{\rm max}\rangle$.
The only one point of notice is that we must avoid such $w(\btau)$ as effectively apply a strong staggered field that conflicts the anisotropy.


\section{Results}\label{sec:2DXXZresults}

\subsection{The unfrustrated region ($j_d=0$) }\label{subsec:unfru}

First, the VDMA calculation has been performed for the 2D XXZ model on the square lattice ($j_d=0$) . This model has been studied intensively  with several numerical methods and thus reliable data are available. Hence we here compare the VDMA results with those already-known data to investigate the numerical accuracy of the VDMA.

\begin{figure}
\includegraphics[width=7cm,clip]{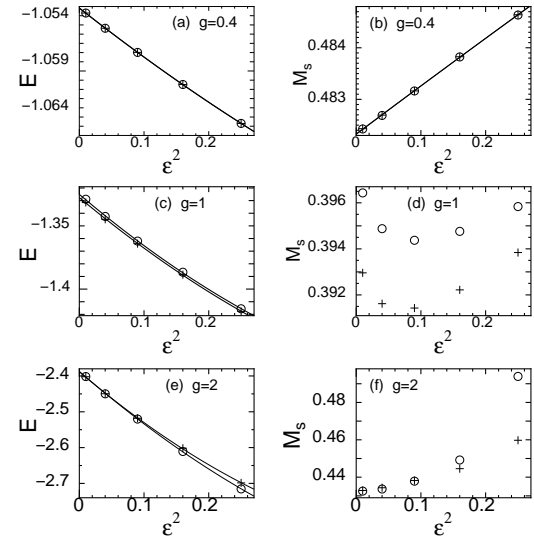}
\caption{ Trotter extrapolation of the energy $E$ and the staggered magnetization $M_s$. Open circles denote results for $(M,m)$=$(2,8)$, and crosses for $(3,9)$. The solid lines are fits of data.}
\label{fig:eext} 
\end{figure}

Figure~\ref{fig:eext} shows the $\epsilon$-dependence of the ground state energy per plaquette $E$ and the staggered magnetization per site $M_s$ for $(M,m)$=$(2,8)$ (open circles) and $(3,9)$ (crosses). Here $M_s$ is the $z$-component ($M_s^z$) of the staggered magnetization for the Ising-like model $g<1$, the $x$-component ($M_s^x$) for the XY-like model $g>1$, or the absolute value ($\sqrt{(M_s^x)^2 + (M_s^z)^2}$) for the Heisenberg model $g=1$. It has been confirmed that the convergence with respect to $m$ is sufficient in the whole $g$ region. 

When using numerical methods based on the Suzuki-Trotter decomposition, the extrapolation for $\epsilon \to 0$ should be performed to obtain the true values of physical quantities. Then the following function
\begin{equation}
O(\epsilon) = O(\epsilon=0) + a_2\epsilon^2 + a_4\epsilon^4, \label{eq:ene-extra}
\end{equation}
where $a_2$ and $a_4$ are constants, is commonly employed for an observable $O$.~\cite{trotter-extra} The $E$-$\epsilon^2$ plot always shows the nearly linear function (see Fig.~\ref{fig:eext} (a,c,e) ). Thus eq.~(\ref{eq:ene-extra}) can be used for extrapolation of the energy in the whole $g$ region. For the staggered magnetization $M_s$, in the Ising-like region ($g < 0.6$) $M_s$-$\epsilon^2$ plot also shows almost linear dependence. The $g = 0.4$ case is illustrated in Fig.~\ref{fig:eext} (b). By contrast, for the (nearly) critical systems ($g \ge 0.6$), the almost linear dependence cannot be observed, which is attributed to the insufficient $M$-convergence for each $\epsilon$ (see Fig.~\ref{fig:eext} (d),(f)). Then $M_s$ at $\epsilon=0.1$ are used as a rough estimate of $M_s(0)$.

\begin{table}
\caption{ Results of the ground state energy $E$ and the staggered magnetization $M_s$.}
\begin{tabular}{lccc}
\multicolumn{4}{c}{ $E$ } \\ \hline
g      & ($M,m$)=($2,8$)   &  ($M,m$)=($3,9$) & QMC \\ \hline
0.4    &  -1.0532    &  -1.0532    & -1.056 ~\cite{barnes} \\
0.8    &  -1.2106     &  -1.2110    & -1.214 ~\cite{barnes} \\
1.0    &  -1.3250     &  -1.3276    & -1.33887 ~\cite{sandvik} \\
1.2    &  -1.5314     &  -1.5329    & -1.538 ~\cite{barnes} \\
2.0    &  -2.3871     &  -2.3884    &  -2.4  ~\cite{okabe} \\ \hline \hline
\multicolumn{4}{c}{ $M_s$ } \\ \hline
g      & ($M,m$)=($2,8$)   &  ($M,m$)=($3,9$) & QMC \\ \hline
0.4    &  0.482     &  0.482     &  0.48 ~\cite{barnes} \\
0.8    &  0.432     &  0.431    &  0.41 ~\cite{barnes} \\
1.0    &  0.396     &  0.393    &  0.3070 ~\cite{sandvik} \\
2.0    &  0.433     &  0.433    &  0.41   ~\cite{miyashita} \\
\end{tabular}
\label{tab:table2}
\end{table}

In Table~\ref{tab:table2}, results of $E$ and $M_s$ are compared with several QMC results~\cite{okabe,miyashita,barnes,sandvik}. The relative deviations from the QMC results are the order of $1\%$.
For $M_s$, the quantitative agreement with the QMC results is not so good as that of $E$; at the isotropic point, the VDMA result of $M_s$ is about $30\%$ larger than a recent QMC result.~\cite{sandvik}  However, it has been confirmed that the VDMA results can reproduce the qualitative behavior of $M_s$, such as the phase transition between the Ising-like N\'{e}el phase ($M_s^z \ne 0, M_s^x=0$) and the XY-like N\'{e}el phase ($M_s^z=0, M_s^x\ne 0$) at $g=1$.

Here, the author gives a comment on the calculation time of the VDMA. The dominant part of the VDMA calculation is the implementation of the DMRG (or CTMRG) for the 2D classical system, and the calculation time of this part scales as $m^2 \times M^8$. Hence it is highly sensitive to $M$. For example, one VDMA calculation with $(M,m)=(3,9)$ takes dozens of hours of workstation time on an Alpha 21264 workstation, although that with $(M,m)=(2,8)$ takes only one or two hours. Thus it is unreasonable to further increase $M$.

\subsection{The frustrated region ($j_d\ne 0$)}

In this subsection  the VDMA results for the frustrated region ($j_d \ne 0$) are presented. The most important topic in this region is the presence of a magnetically disordered phase at the ``2D pyrochlore'' point ($j_d=g=1$).~\cite{palmer,canals,elhajal,moessner,fouet,brenig,brenig2,starykh,sind,auer} In addition, in the isotropic case ($g=1$) the disordered phase is considered to survive up to $j_d=0.75$.~\cite{canals,sind} In the following the chief object of interest is the system in the anisotropic region ($g\ne 1$), which has not been explored. 

\begin{figure}
\includegraphics[width=6cm,clip]{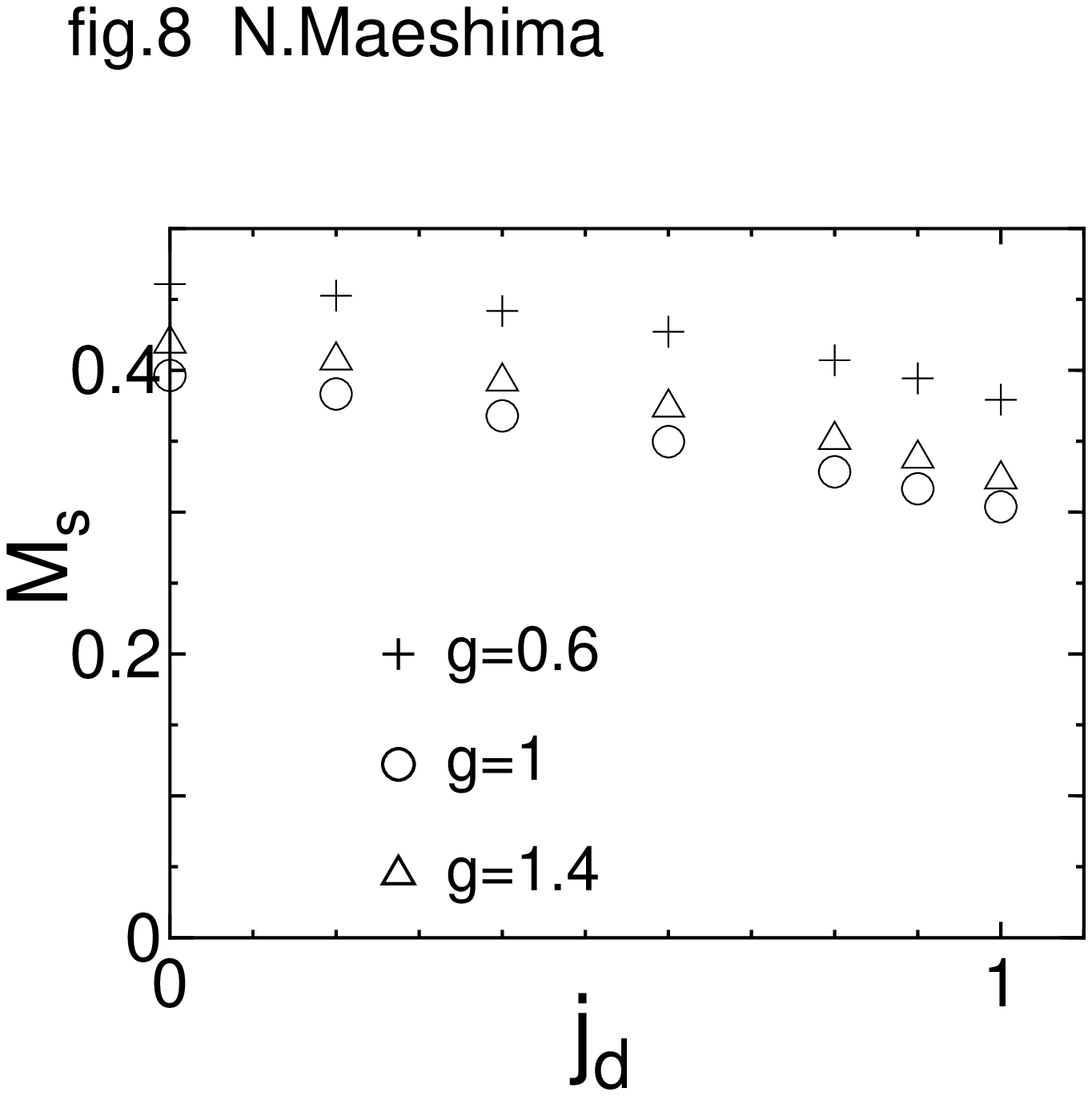}
\caption{The staggered magnetization for finite $j_d$.}
\label{fig:mag-check}
\end{figure}

In Fig.~\ref{fig:mag-check}, the staggered magnetization $M_s$ is plotted as a function of $j_d$ for the Ising-like system ($g=0.4$), the isotropic system ($g=1$), and the XY-like system ($g=1.4$). VDMA results shown in this subsection are obtained by calculations with $(M,m)$=$(2,16)$.  The extrapolation procedure for $\epsilon \to 0$ is same as that of the unfrustrated case. The obtained results show that the finite staggered magnetization $M_s$ remains in the range $0\le j_d \le 1$ for all $g$. For $g=1$, this is contradict with the observation of the magnetically disordered phase for $j_d\ge 0.75$.~\cite{canals,sind}

The disordered phase observed at the 2D pyrochlore point is called the valence bond crystal (VBC),~\cite{moessner,fouet,brenig,brenig2} where the array of 4 spin singlets located on the void squares is the basic structure of the ground state. Here  a simple argument is given based on the VBC picture to compare with the VDMA results. For this purpose, the VBC state is generalized for the anisotropic system. At the isotropic point the pure VBC state is a direct product of the ground state of a Heisenberg chain with only 4 spins. Hence for $g \ne 1$ the generalized VBC state $|\psi_{VBC}\rangle$ can be given as a direct product state of the ground states of the 4 spin chain with the XXZ-type interaction. Then the expectation value of the energy ($E_{VBC}=-1/4-(\sqrt{1/4+2g^2})/2$) can be obtained, which is an upper bound of the ground state energy.
$E_{VBC}$ is independent of $j_d$ because of vanishing correlation of the next-nearest neighbor pairs interacting with $j_d$.

\begin{figure}
\includegraphics[width=6cm,clip]{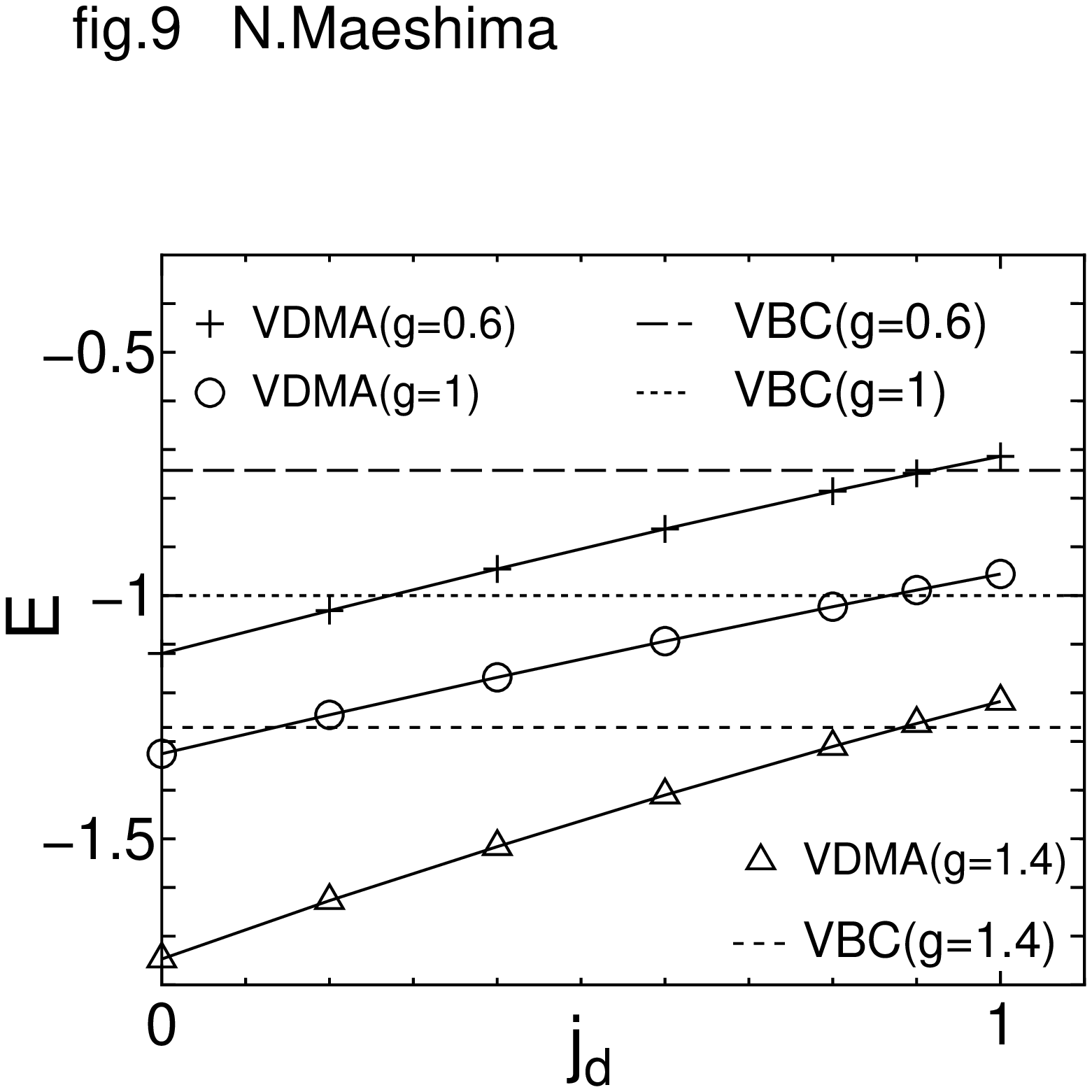}
\caption{The ground state energy  $E$. Symbols denote the VDMA results and the solid lines are guides to the eye. The dashed, dotted, and short-dashed lines show $E_{VBC}$.}
\label{fig:ene-check}
\end{figure}

Figure~\ref{fig:ene-check} shows the VDMA results of the ground state energy $E$ and the upper bound $E_{VBC}$. It is found that the VDMA results for $E$ is lower than $E_{VBC}$ in the region where the frustration effect is not strong. However, the VDMA result for $E$ exceeds $E_{VBC}$ around $j_d=1$, which implies that the VDMA calculation does not reach the true ground state there. This is the reason why the VDMA gives  finite $M_s$  even near $j_d=1$ for $g=1$.


\begin{figure}
\includegraphics[width=6cm,clip]{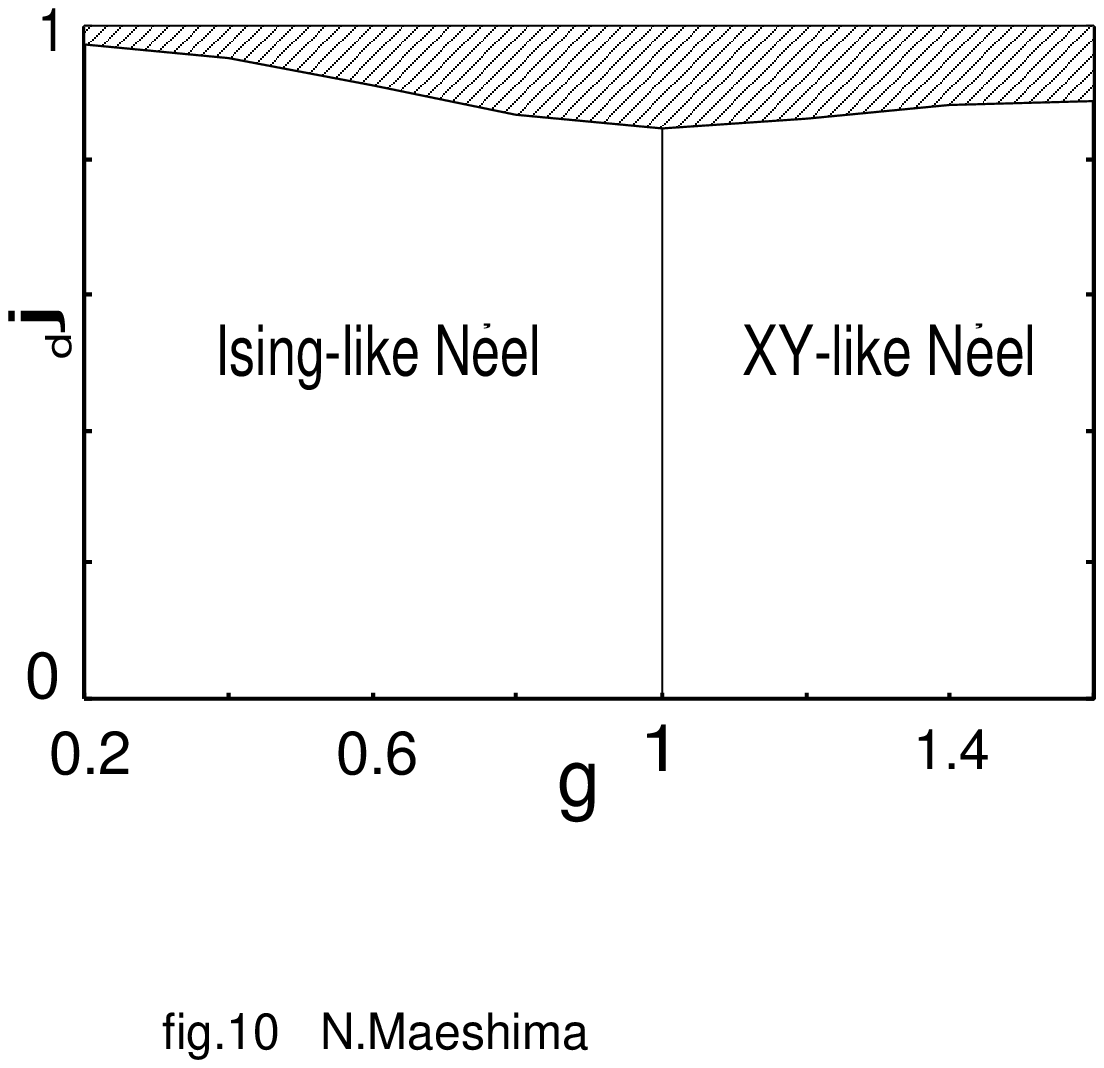}
\caption{Phase diagram of the S=1/2 XXZ model on the checkerboard lattice.
In the shaded region, the generalized VBC state gives lower energy than the VDMA result.}
\label{fig:boundary}
\end{figure}

The comparison between the VDMA results and the upper bound $E_{VBC}$ leads to the phase diagram depicted in Fig.~\ref{fig:boundary}.
 Only from the VDMA, it is concluded that the whole parameter region is in the N\'{e}el phase. The $g=1$ line is the phase boundary between Ising-like phase and the XY-like one.
Although the presence of the N\'{e}el phase is valid in the weakly frustrated region, 
it is quite doubtful in the shaded region in Fig.~\ref{fig:boundary}, where the generalized VBC state gives lower energy than the VDMA result.
The remaining problem is what is the realized phase in the shaded region. Naturally, the VBC phase is the prime candidate.
In particular, it is likely that the VBC phase extends around the $g=1$ line within $0.75 \le j_d \le 1$.~\cite{canals,sind} However, it is unclear whether the VBC phase is realized far away from the 2D pyrochlore point. 
Hence, at present, we cannot reach a definite conclusion on the shaded region. Thus, further theoretical investigation is needed to complete the ground state phase diagram of this model.

\begin{figure}
\includegraphics[width=5cm,clip]{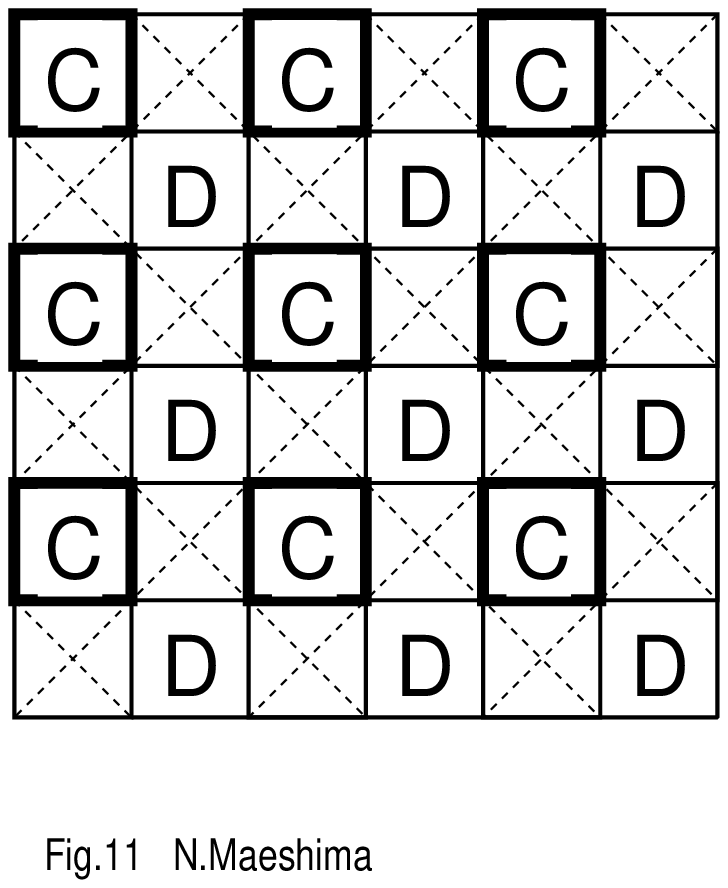}
\caption{The valence bond crystal on the checkerboard lattice. The bold squares show the four-spin singlets.}
\label{fig:vbc}
\end{figure}

Finally, let us consider how we can obtain a good approximation of the true ground state in the shaded region of Fig.~\ref{fig:boundary}.
One possible solution in the TPS formulation is to construct a new variational TPS that can represent the pure VBC state.
The simplest one of such a state can be  made as follows.
As shown in ref.~\cite{fouet}, the pure VBC state has the staggered pattern of the 4 spin singlets on the void squares (see Fig.~\ref{fig:vbc}). Hence the void squares can be divided into two groups C and D;
the group C corresponds the void squares that have the singlet states
and group D corresponds those in which 4 spins are uncorrelated.
This structure of the VBC needs two types of tensors in the TPS representation; one constructs
the 4 spin singlet and another represents the uncorrelated 4 spin state.
Thus the TPS form of the pure VBC state is denoted as
\begin{equation}
\langle[\bsigma]|VBC\rangle = \prod_{s(ij)\in C}X_C(\bsigma_{ij}) \prod_{s(ij)\in D}X_D(\bsigma_{ij}) 
\end{equation}
and the two types of tensors are given as
\begin{eqnarray}
 X_C(\sigma_1,\sigma_2,\sigma_3,\sigma_4) 
&=& \left\{ \begin{array}{@{\,}ll} 1 & \mbox{for $(\sigma_1,\sigma_2,\sigma_3,\sigma_4)$}\\
           & \mbox{= ($\uparrow\downarrow\uparrow\downarrow$) or ($\downarrow\uparrow\downarrow\uparrow$) } \\ & \\
 -\frac{1}{2} & \mbox{for $(\sigma_1,\sigma_2,\sigma_3,\sigma_4)$} \\
  & \mbox{= ($\uparrow\uparrow\downarrow\downarrow$),($\uparrow\downarrow\downarrow\uparrow$)} \\
  & \mbox{($\downarrow\downarrow\uparrow\uparrow$),or ($\downarrow\uparrow\uparrow\downarrow$)} \\
  & \\
  0 & \mbox{for other} \\
    & (\sigma_1,\sigma_2,\sigma_3,\sigma_4),
\end{array}  \right. 
\end{eqnarray}
and
\begin{equation}
 X_D(\sigma_1,\sigma_2,\sigma_3,\sigma_4) = 1 \mbox{ for any $(\sigma_1,\sigma_2,\sigma_3,\sigma_4)$}.
\end{equation}
Then a straightforward variational scheme based on the above argument is to take the several (or all)
 components of $X_C$ and $X_D$ as variational parameters and to find the minimum of the variational energy with varying those parameters.

It should be noted that such a variational TPS is the simplest one to represent the VBC state. The tensors $X_C$ and $X_D$ have only the bare spin variables $\bsigma$, which corresponds to the case of $M=1$.
Hence more complicated TPS having auxiliary variables with $M\ge 2$ would give lower variational energy because  it has more variational parameters.

The above discussion also tells us the reason why we have failed to reach the VBC state.
In this study, all the tensors are put on the plaquettes that has the cross bond, not on the void plaquette. Hence the TPS used here does not have the natural form to represent the pure VBC state.
Thus, to complete the phase diagram, we probably need to use the TPS that reflects the spatial distribution of the 4 spin singlets in the VBC state.

\section{Conclusion}
In this paper the vertical density matrix algorithm (VDMA) is applied to the spin $1/2$ XXZ model on the checkerboard lattice. The VDMA for this model needs several modifications in its implementation procedure because of the lattice structure of the 3D classical spin system mapped by the Suzuki-Trotter transformation.

The ground state energy and the staggered magnetization are calculated for the unfrustrated model to investigate the numerical accuracy of the VDMA. The obtained data show that the ground state energy is quantitatively reliable; the maximum deviation from the QMC results is about $1\%$ for the unfrustrated Heisenberg model. For the staggered magnetization the VDMA only reproduces the qualitative behavior of the 2D XXZ model.

In the frustrated region, the VDMA results are compared with the simple calculation based on the VBC state, which is considered to be realized at the 2D pyrochlore point. It is concluded that, in the weakly frustrated region, the system is in the N\'{e}el ordered phase.
By contrast, in the strongly frustrated region where the VDMA does not reach the true ground state, the realized phase cannot be identified. Hence to complete the phase diagram of this model, we need more reliable data for the strongly frustrated region in the thermodynamic limit. For this purpose, further improvement of the VDMA or a novel tensor product formulation would be useful.

\section*{Acknowledgments}
The author would like to thank Professor Y. Akutsu, Dr. K.~Okunishi, Dr. T.~Nishino, and Dr. Y.~Hieida for continuous discussions and comment on the TPS formulations for higher dimensional systems. He also thanks Professor M.~Kikuchi for valuable comments. This study was partially supported
by IT-program of Ministry of Education, Culture, Sports, Science and Technology.

\end{document}